\documentclass[journal,twoside,web]{ieeecolor}
\usepackage{generic}
\usepackage{cite}
\usepackage{amsmath,amssymb,amsfonts}
\usepackage{algorithmic}
\usepackage{graphicx}
\usepackage{algorithm,algorithmic}
\usepackage{hyperref}
\hypersetup{hidelinks=true}
\usepackage{textcomp}
\def\BibTeX{{\rm B\kern-.05em{\sc i\kern-.025em b}\kern-.08em
    T\kern-.1667em\lower.7ex\hbox{E}\kern-.125emX}}
\begin{document}

\title{Non-Markovian Dynamical Systems Modeling of Electroencephalogram-based Brain Activity for Predicting the Cognitive Fatigue Level}
\author{Zeinabsadat Saghi, Daria Riabukhina, Olubukola Akinbami, Paul Bogdan, Souti Chattopadhyay 
\thanks{\textbf{This paper is submitted for review at IEEE Journal of Biomedical and Health Informatics (JBHI).}}
\thanks{Zeinabsadat Saghi is with Thomas Lord Department of Computer Science, University of Southern California, Los Angeles, USA (e-mail: saghi@usc.edu)}
\thanks{Daria Riabukhina, Ming Hsieh Department of Electrical and Computer Engineering, University of Southern California, Los Angeles, USA (e-mail: riabukhi@usc.edu)}
\thanks{Olubukola Akinbami is with Weinberg Institute for Cognitive Science, University of Michigan, Ann Arbor, USA (e-mail: oaknbami@umich.edu)}
\thanks{Paul Bogdan is with Ming Hsieh Department of Electrical and Computer Engineering, University of Southern California, Los Angeles, USA (e-mail: pbogdan@usc.edu)}
\thanks{Souti Chattopadhyay is with Thomas Lord Department of Computer Science, University of Southern California, Los Angeles, USA (e-mail:schattop@usc.edu) } 
}

% \begin{affiliations}
% Z. Saghi, S. Chattopadhyay\\
% Thomas Lord Department of Computer Science, University of Southern California, Los Angeles, USA

% D. Riabukhina, P. Bogdan\\
% Ming Hsieh Department of Electrical and Computer Engineering, University of Southern California, Los Angeles, USA\\
% pbogdan@usc.edu

% O. Akinbami\\
% Weinberg Institute for Cognitive Science, University of Michigan, Ann Arbor, USA

% \end{affiliations}

% \thanks{This paragraph of the first footnote will contain the date on 
% which you submitted your paper for review. It will also contain support 
% information, including sponsor and financial support acknowledgment. For 
% example, ``This work was supported in part by the U.S. Department of 
% Commerce under Grant 123456.'' }
% \thanks{The next few paragraphs should contain 
% the authors' current affiliations, including current address and email. For 
% example, First A. Author is with the National Institute of Standards and 
% Technology, Boulder, CO 80305 USA (email: author@boulder.nist.gov). }
% \thanks{Second B. Author Jr. was with Rice University, Houston, TX 77005 USA. He is 
% now with the Department of Physics, Colorado State University, Fort Collins, 
% CO 80523 USA (email: author@lamar.colostate.edu).}
% \thanks{Third C. Author is with 
% the Electrical Engineering Department, University of Colorado, Boulder, CO 
% 80309 USA, on leave from the National Research Institute for Metals, 
% Tsukuba, Japan (email: author@nrim.go.jp).}}

\maketitle

\begin{abstract}
Cognitive fatigue transitions from focused attention to inexact responses can cause catastrophic failures in high-stakes environments, yet current black-box assessment techniques ignore the brain's non-Markovian and time-varying interdependent properties, limiting real-time phase transition detection. We develop a fractional dynamical networks-based machine learning (FDNML) framework using coupled fractional-order differential equations to capture brain signal interdependencies and detect cognitive fatigue transitions in real-time. Multifractal properties of brain activity exhibit distinct generalized fractal dimension signatures across fatigue levels, with Wasserstein distances of 0.10, 0.13, and 0.08 between states 0-1, 1-2, and 0-2, respectively. The framework achieves 93.33\% classification accuracy and 95\% AUROC, enabling prevention of performance degradation through early detection of neural state transitions. 
\end{abstract}

\begin{IEEEkeywords}
Cognitive Fatigue, Complexity Analysis, Multifractal Analysis
\end{IEEEkeywords}

\section{Introduction}

% \cite{wiehler2022neuro} this paper modeled cognitive fatigue as a neuro-metabolic model/process that is as the cost of cognitive control.

Cognitive fatigue manifests as a continuous transition from a state of focused attention and agile decision making to a state of inexact and slow response~\cite{wiehler2022neuro,wang2016compensatory,borragan2017cognitive}. Although often underestimated, the emergence of cognitive fatigue and its consequences can be catastrophic---particularly in high-stakes domains as seen repeatedly in history during the Chernobyl nuclear disaster~\cite{Mitler1988CatastrophesSA}, Exxon Valdez oil spill~\cite{harrington2001health,techera2016causes}, Challenger space shuttle explosion~\cite{Mitler1988CatastrophesSA}. Existing cognitive fatigue assessment techniques~\cite{karim2024examining} focus on black box models to mine large amounts of brain data and mostly perform only binary classification of fatigue, which limits their utility for clinical research where fatigue evolves continuously and must be tracked sensitively over time~\cite{Martins2021Fatigue}.
However, these machine learning techniques ignore the intrinsic non-Markovian and time-varying interdependent properties of the brain in action, preventing agile detection of fatigue's phase transitions.

Avoiding states of poor cognitive performance that can lead to catastrophic events requires novel, accurate mathematical models and algorithmic strategies that enable us to mine the brain activity in real time and answer the following questions: \emph{How do brain signal patterns change with increasing cognitive fatigue levels?} \emph{How do the different cognitive fatigue levels impact the brain's network dynamics?} \emph{Can we detect cognitive fatigue phase transitions in real time?}

To address these questions, we develop an interpretable fractional dynamical networks-based machine learning (FDNML) framework capable of (1) mining the brain activity in real-time and capturing the non-Markovian and time-varying interdependence among a set of brain signals through a set of coupled fractional order differential equations, (2) detecting the real-time transitions from one cognitive level to another. The FDNML framework shows that the multifractal properties of brain activity change across various cognitive fatigue levels, observable through distinct generalized fractal dimension (GFD) $D_q$ plots for each level ($\mu$ Wasserstein distance between fatigue levels 0-1: 0.10, 1-2: 0.13, 0-2:0.08).

% ) analyzing the network complexity of the cognitive fatigue as it unfolds, and (3
Our framework also identifies that the complexity of the interdependence of brain signals changes from one cognitive fatigue level to another. Finally, our model is able to classify cognitive fatigue levels with an accuracy of 93.33\% and an AUROC of 95\% using deep learning models.

\section{Related Works}
Recent studies highlight the potential of multifractal analysis in evaluating cognitive functions through EEG data \cite{gaurav_anand_kumar_2021}and \cite{tibarewala_2017}. This approach overcomes the limitations of traditional assessments, objectively capturing complex neural patterns linked to cognitive fatigue \cite{gaurav_anand_kumar_2021}. As cognitive stress intensifies, these patterns often shift from multifractal to monofractal dynamics, revealing indicators of mental overload \cite{tibarewala_2017}. Cho et al. explore the application of fractal features, specifically the use of asymmetric Hurst exponents, to enhance pattern recognition and prediction capabilities \cite{cho_lee_2022}. With these insights, researchers improve the accuracy of detecting changes in cognitive workload, playing a critical role in developing interventions that effectively manage cognitive fatigue.

\subsection*{Multifractal Features and Cognitive Fatigue}
% (multifractal analysis only)
In the realm of cognitive fatigue, multifractal analysis using the generalized Hurst exponent (Dq) provides valuable insights\cite{gaurav_anand_kumar_2021}. Measuring Dq allows researchers to observe changes in brain efficiency during cognitive tasks, indicating levels of fatigue\cite{tibarewala_2017}. This novel approach accurately monitors cognitive resource usage and helps identify mental fatigue \cite{gaurav_anand_kumar_2021}. Our research uniquely leverages these features to establish a strong connection between the complexity of brain signals and cognitive fatigue, paving the way for enhanced detection methods that recognize fatigue and changes in cognitive performance.
\subsection*{Multifractal and Neural Network Pipeline}
% (Cognitive Fatigue and Neural networks- works where they analyze raw data and use neural networks)
This research introduces a novel approach that integrates multifractal features with neural networks like Long Short-Term Memory (LSTM) and Deep Neural Networks (DNN) to analyze EEG data \cite{gaurav_anand_kumar_2021}. The method utilizes the temporal dynamics captured by Dq, allowing for accurate predictions of cognitive performance and fatigue \cite{tibarewala_2017}. By employing these advanced neural models, we streamline real-time cognitive monitoring processes, demonstrating potential impacts where maintaining mental efficiency and adaptive safety responses becomes critical. This research showcases the effectiveness of a combined approach, setting new standards in cognitive performance evaluation \cite{tibarewala_2017}. In \cite{karim2023fatigue}, the authors compare 1D-CNN, RNN, LSTM, and EEGNet for the detection of induced cognitive fatigue. EEGNet significantly outperforms other models with 88.17\% accuracy, proving that carefully designed neural networks are good at detecting cognitive fatigue. In \cite{gao2023logmelcrnn}, Gao et al. transform drivers' fatigue into log-Mel diagrams and then use them as input for a convolutional-recurrent neural network. This approach allows for capturing both spectral-spatial and long-range time dependencies.The performance of the model exceeds 88\% accuracy, outperforming existing methods based on CNN and LSTM. The results can be used in practice for real-time driver fatigue detection. The study \cite{lee2024pilotfatigue} considers multichannel EEG of ten pilots in a flight simulator and classifies the results as normal, low fatigue, and high fatigue. The proposed model, consisting of 5 convolutional layers and one LSTM layer, outperforms baseline models with 88.01\% average accuracy across all participants. The work is valuable for pilot fatigue monitoring in real time, which is crucial for safe aviation.

\section{Method}
\subsection{Multifractal analysis of EEG signals}

Multifractal analysis is a statistical method that captures variations in signal regularity across scales\cite{doukhan2002theory}. It is well-suited for characterizing complex, non-uniform behaviors in time series data such as EEG. This section outlines the analytical approach and statistical procedures used. To examine scale-invariant properties, we estimate the local regularity of the signal using the H\"older exponent \( h(t_0) \), which measures smoothness at each time point. Higher values indicate smoother activity, while lower values reflect more irregular changes. The distribution of these exponents is summarized by the multifractal spectrum \( D(h) \), which describes how regularity varies over the signal.
We estimate \( D(h) \) using the wavelet leader multifractal formalism. This method relies on the discrete wavelet transform (DWT) to compute wavelet coefficients \( d_X(j, k) \), which represent the signal at different scales. From these, we derive wavelet leaders \( L_X(j, k) \), defined as the maximum wavelet coefficient within a local neighborhood across finer scales. These leaders accurately track local regularity.

To quantify scaling behavior, we compute structure functions \( S_L(j, q) \), defined as the average \( q \)-th power of wavelet leaders at scale \( j \):
\begin{equation}
S_L(j, q) = \frac{1}{n_j} \sum_{k=1}^{n_j} L_X(j, k)^q.
\end{equation}

Structure functions follow a power-law form,
\begin{equation}
S_L(j, q) \sim 2^{j\zeta(q)},
\end{equation}
where \( \zeta(q) \) are scaling exponents estimated via linear regression in log-log space. We derive the multifractal spectrum \( D(h) \) by applying a Legendre transform:
\begin{equation}
D(h) = \min_{q \neq 0} (1 + qh - \zeta(q)).
\end{equation}

To simplify interpretation, we compute log cumulants \( c_p \) from the logarithm of wavelet leaders. The first cumulant \( c_1 \) gives the peak of \( D(h) \), \( c_2 \) indicates its width (range of exponents), and \( c_3 \) reflects asymmetry. These values summarize the multifractal profile concisely and are useful for classification tasks. We also assess whether a signal is monofractal or multifractal. Monofractal signals have a single exponent and linear \( \zeta(q) \), i.e.,
\begin{equation}
\zeta(q) = qH.
\end{equation}
Multifractal signals show nonlinear \( \zeta(q) \) and nonzero \( c_2 \). We use \( c_2 \) to detect multifractality. To ensure robust inference, we apply a block bootstrap procedure. This resampling method preserves the temporal structure of the signal and allows us to compute confidence intervals and \( p \)-values without assuming Gaussianity. Bootstrap results support the reliability of estimated exponents and cumulants.

\subsection{Fractional Dynamical Networks-based Machine Learning}

In this section, we explain our model structure, as shown in ~\ref{fig:overview}. 

Let \( X_r \in \mathbb{R}^{n \times d_r} \) be the raw EEG signal and \( X_f \in \mathbb{R}^{n \times d_f} \) be the feature matrix, where \( n \) is the number of samples, \( d_r \) and \( d_f \) are the respective dimensionalities.

The parallel encoders are defined as:
\begin{align*}
f_r: X_r &\to \mathbb{R}^{n \times k} \\
f_f: X_f &\to \mathbb{R}^{n \times k}
\end{align*}
where \( k \) is the embedding dimension. The contrastive loss function is formulated as:

\begin{equation}
\begin{split}
\mathcal{L}_{\text{contrastive}}(f_r, f_f) = -\frac{1}{N} \sum_{i=1}^{N} \Bigg[ \frac{\text{sim}(z_r^i, z_f^i)}{\tau} \\
- \log \sum_{j=1}^{N} \mathbb{1}_{[j \neq i]} \exp \left( \frac{\text{sim}(z_r^i, z_f^j)}{\tau} \right) \Bigg]
\end{split}
\end{equation}

Where $z_r^i = f_r(x_r^i)$ and $z_f^i = f_f(x_f^i)$ are L2-normalized embeddings, specifically $\hat{z}_r^i = \frac{z_r^i}{\|z_r^i\|_2}$ and $\hat{z}_f^i = \frac{z_f^i}{\|z_f^i\|_2}$. The similarity function $\text{sim}(a,b)$ is the dot product $a^\top b$ (which is equivalent to cosine similarity for unit vectors). The temperature parameter is set to $\tau = 0.2$, $N$ is the batch size.

The objective is to minimize the following loss, which encourages:
\begin{enumerate}
\item Positive pair alignment by maximizing \( \hat{z}_r^i \cdot \hat{z}_f^i \)
\item Negative pair uniformity by minimizing \( \hat{z}_r^i \cdot \hat{z}_f^j \) for \( i \neq j \)
\end{enumerate}

\noindent The symmetric implementation computes:
\begin{equation}
\mathcal{L} = \frac{1}{2N} \sum_{i=1}^N \left[ \mathcal{L}(z_r^i, z_f^i) + \mathcal{L}(z_f^i, z_r^i) \right].
\end{equation}

\subsection{Complexity Index Analysis}

To calculate the CI, we first convert all sample data—represented in our dataset as a 202 $\times$ 16 coupling matrix—into a binary sequence using the median value as a threshold. $\mathbf{X} \in \mathbb{R}^{202 \times 16}$. We vectorize the data with $\phi: \mathbb{R}^{202 \times 16} \rightarrow \mathbb{R}^n$, where $n = 202 \times 16 = 3232$, so that $\mathbf{x} = \phi(\mathbf{X}) = (x_1, \dots , x_n)$. We define the sequence $s_i = \mathbf{1}\{x_i > m\} \in \{0,1\}$, $i= 1,\dots,n$, where $m=median(\mathbf{x})$. 

The asymptotic complexity value for a binary sequence is:
\begin{equation}
c(n) \sim \frac{n}{\log_2(n)}.
\end{equation}

The normalized LZC is calculated as:
\begin{equation}
\text{CI}_\text{LZC} = \frac{c}{n / \log_2(n)}.
\end{equation}

In general,
\begin{equation}
\lim_{n \to \infty} \frac{c}{n / \log_2(n)} = H,
\end{equation}
where $H$ is the entropy rate in bits/symbol \cite{lempel_ziv}.

\section{Results}
% \subsection{CogBeacon dataset}

% Participants indicated their fatigue level by pressing a button while performing tasks, with each level representing the cumulative fatigue reported up to that turn. Reported fatigue levels ranged from 0 (no fatigue) to 6 (maximum fatigue).

% Due to the limited number of participants reporting higher fatigue levels (4, 5, 6), our analysis focuses on fatigue levels 1, 2, and 3, which were more prevalent among participants.
\begin{figure*}[h!]
    \centering
    \includegraphics[width=\linewidth]{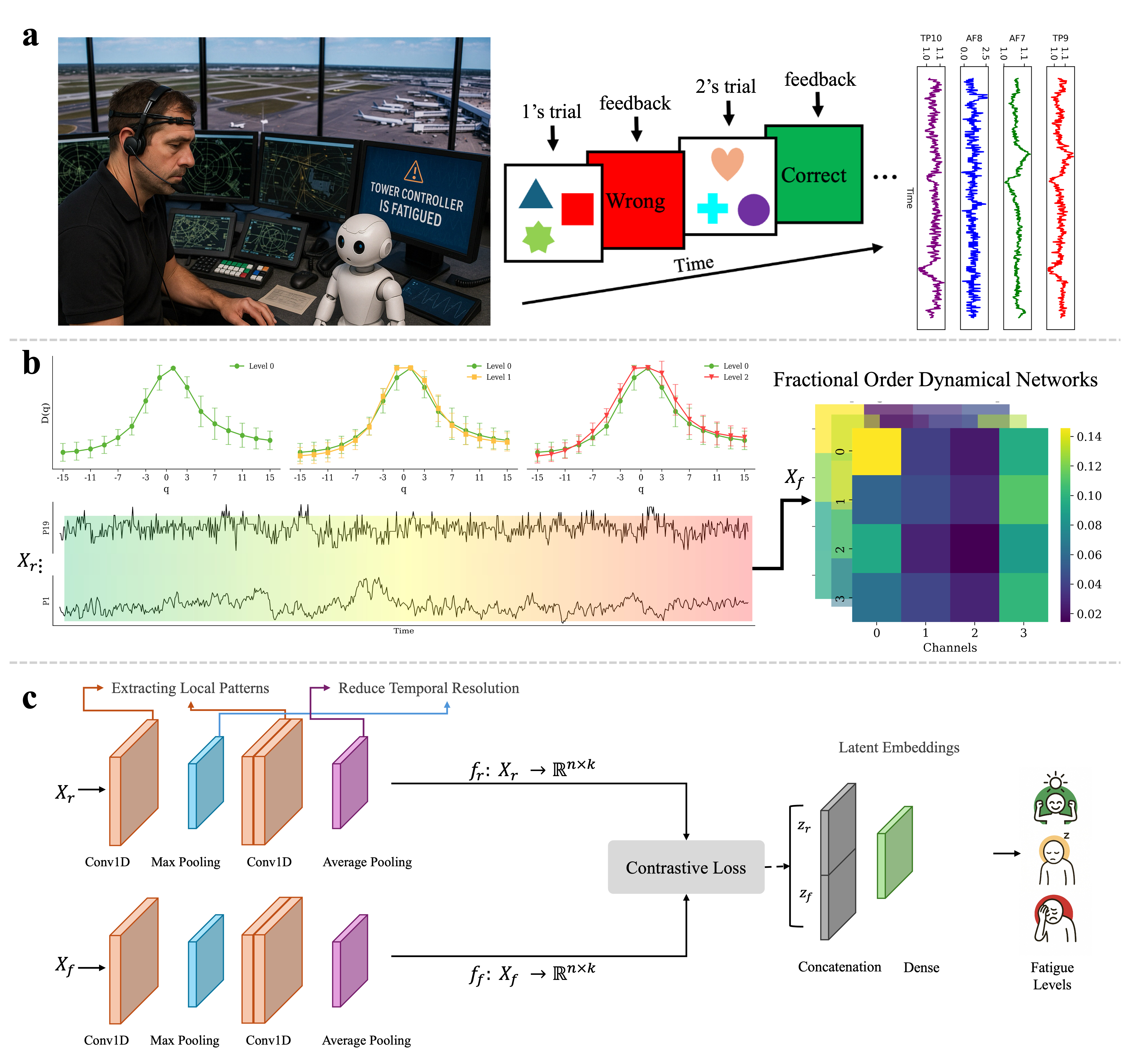}
    \caption{\textbf{Overview of the proposed fractional dynamical networks-based machine learning framework for cognitive fatigue prediction}: a) Schematic representation of an individual performing the Wisconsin cognitive task (middle image) and its recording of four EEG channels reflecting their brain activity and various cognitive fatigue levels. b) Multifractal analysis of the AF7 EEG channel of one participant shows that the generalized fractal dimension changes as the level of cognitive fatigue increases. For a comprehensive understanding of cognitive fatigue, our FDNML framework estimates the time-varying networks (i.e., coupling matrices) among the EEG signals, allowing us to quantify their differences and complexity. c) A fractional dynamical network modeling of brain activity provides essential features about the multifractality and time-varying interdependence among EEG channels for the real-time machine learning prediction task.}
    \label{fig:overview}
\end{figure*}

\vspace{1em}
\noindent\textbf{The transition from high cognitive performance to varying degrees of fatigue is marked by distinct multifractal patterns in brain activity.} While numerous analyses inspired by nonlinear dynamics and statistical physics have highlighted the fractal patterns and scale invariance properties of physiological processes (e.g., EEG~\cite{Peng2002QuantifyingFD,Chang2007FractalDE,Schumann2010AgingEO}, heart rate~\cite{ivanov1999multifractality,goldberger2002fractal}, blood glucose~\cite{ghorbani2013cyber}), no comprehensive analysis investigated how cognitive fatigue manifests itself across large networks of nonlinear dynamical systems that encompass brain activity. While the fractality of a signal implies scale-invariance behavior, meaning that its statistical properties remain consistent regardless of the observation scale, the multifractality of the signal refers to a generalized scaling behavior where the fractal scaling exponent varies across different time scales. Moreover, the multifractal analysis of various signals measured from a complex dynamical system provides information not only about a continuous spectrum of fractal exponents but also about the system's entropy and free energy. Adopting this non-equilibrium statistical physics perspective, we hypothesize that the cognitive fatigue of the brain in action and in context manifests itself as a phase transition from a state of focused attention and agile fluid intelligence to a state marked by a slow and inexact response. To investigate the existence of such a phase transition, we continuously estimate the generalized fractal dimension (GFD)$D_q$ from each participant's EEG data using a wavelet-based multifractal method~\cite{jaffard2016p,leonarduzzi2016p} during various cognitive tests (see details about the dataset in Methods). While performing various required cognitive tests, each participant reported her/his fatigue level by pressing a button; the frequency of button presses was aggregated for each trial. Figure~\ref{fig:overview}(b) shows the average GFD $D_q$ and their confidence intervals for AF7 EEG signal across all individuals and trials and three phases: no fatigue (\textit{Level 0}), reported fatigue once (\textit{Level 1}), and reported fatigue twice (\textit{Level 2}).

\begin{figure*}
    \centering
    \includegraphics[width=\linewidth]{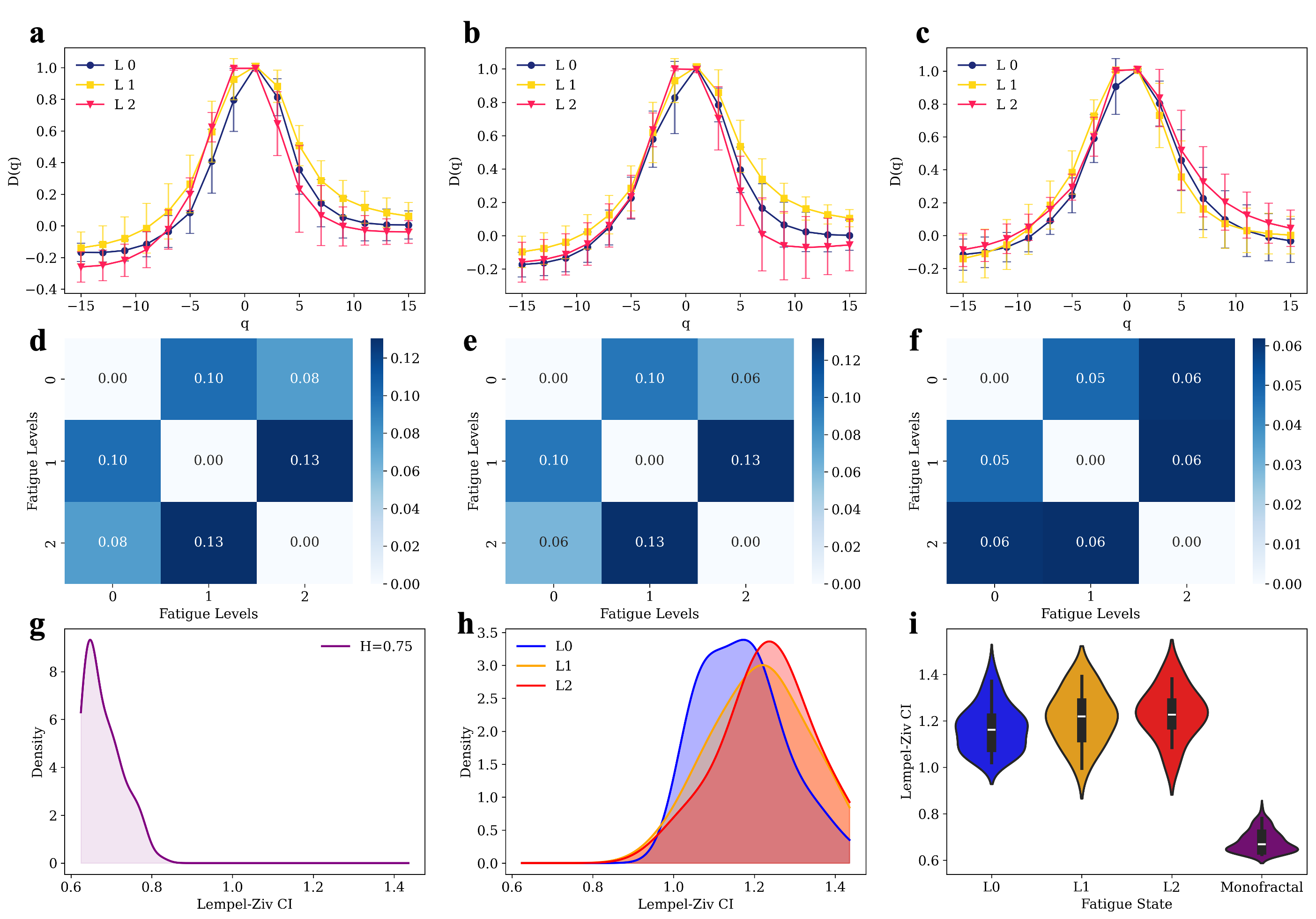}
    \vspace{0.0cm}
    \caption{\textbf{Comparison of Fractal Features across Fatigue Levels with Wasserstein Distance.} \textbf{(a–c)} display the fractal dimension $D_q$ calculated using wavelet leaders across different fatigue levels for three task-channel settings: (a-b) Task V2 for channels TP10 and TP9, and (c) Task V1 for channel TP10. \textbf{(d-f)} illustrate the Wasserstein distance computed for each pair of mean $D_q$ distributions from (a–c), with bars representing the calculated difference between fatigue levels.\textbf{(g-i)} Show the complexity indices calculated for a synthesized monofractal (a) and the average of CI of our generated coupling matrices($A$) of all participants for each fatigue level.}
    \label{fig:ci_dq}
\end{figure*}

% \begin{figure}[htbp]
%     \centering
%     % First figure
%     \begin{subfigure}[b]{0.32\textwidth}
%         \textbf{a} % Add bold label above and aligned to the left
%         \vspace{-1.5em} % Adjust vertical spacing to move the label closer to the image
%         \includegraphics[width=\textwidth]{Figures/spectrum1.png} % Replace with your file name
%     \end{subfigure}
%     % Second figure
%     \begin{subfigure}[b]{0.32\textwidth}
%         \textbf{b} % Add bold label above and aligned to the left
%         \vspace{-1.5em} % Adjust vertical spacing to move the label closer to the image
%         \includegraphics[width=\textwidth]{Figures/spectrum2.png} % Replace with your file name
%     \end{subfigure}
%     % Third figure
%     \begin{subfigure}[b]{0.32\textwidth}
%         \textbf{c} % Add bold label above and aligned to the left
%         \vspace{-1.5em} % Adjust vertical spacing to move the label closer to the image
%         \includegraphics[width=\textwidth]{Figures/spectrum3.png} % Replace with your file name
%     \end{subfigure}
%     \vspace{2em}
%     \caption{\textbf{Comparison of Multifractal Spectrum across fatigue levels.} a–c show changes in the average spectrum from No Fatigue (Level 0) to Higher Fatigue (Levels 1, 2, and 3) across participants.}
% \end{figure}

Figure~\ref{fig:ci_dq} illustrates the analysis of the GFD $D_q$ across the three fatigue levels. GFD analysis demonstrates the existence of variations in the scaling complexity of EEG signals as an individual reports a transition from no fatigue to a more pronounced level of cognitive fatigue. %The plots show how the scaling behavior of EEG dynamics changes with increasing mental fatigue. 
To explore how the cognitive fatigue phase transition manifests, we investigate the curvature (i.e., whether it is narrow or wide) and the range of fractal dimensions $\Delta D_q = D(q_{\text{max}}) - D(q_{\text{min}})$. 
For instance, Figure~\ref{fig:ci_dq}(a) shows the GFD for channel TP10 and $ \Delta D_q $ is more skewed for L1 and L2, while being more symmetric for L0. Figure~\ref{fig:ci_dq}(b) shows the GFD for channel TP9 and follows a similar pattern, but the variation of L2 is higher across participants due to a higher confidence interval error bars in the positive ranges of $q$. Figure~\ref{fig:ci_dq}(d–f) shows the Wasserstein distances between the $ D_q $ curves for each pair of fatigue levels. While the distances are small (ranging between 0.05 and 0.13), the largest differences appear between levels 1 and 2 and levels 0 and 2, respectively. This suggests that fatigue influences the multifractal profile of EEG signals.
% however, the changes are small and difficult to clearly separate.
% The middle row of the figure (d-f) displays Wasserstein distances between the $ D_q $ distributions of each fatigue level pair. While the distances remain modest in absolute value (ranging from approximately 0.06 to 0.13), they are consistently non-zero, especially between L1 and L2 and between L0 and L2. This consistency indicates that mental fatigue does indeed alter the multifractal characteristics of EEG signals, but the differences are subtle and not trivially separable. 
This multifractal analysis validates the existence of distinct fractal properties associated with fatigue states and suggests that modeling such distinctions requires advanced dynamical systems techniques, such as long-range memory-based models, capable of capturing fine-grained temporal structures.

\vspace{2em}
\noindent\textbf{A time-varying fractional dynamical network model can capture the observed non-Markovianity, multifractality, and quantify the complexity of the phase transition from cognitive clarity to cognitive fatigue.} The observed multifractal behavior in Figure \ref{fig:ci_dq}(a-c) (i.e., the GFD varies with the $q$-th order), implies that the brain activity measured by the EEG exhibits specific fractal exponents over predefined time scales. However, based on the level of cognitive involvement of various brain regions on specific tasks, the fractal coefficients may vary from one EEG signal to another. To make the discussion more concrete, we denote by $x_{j}[k] \in \mathbb{R}$ the $j$-th EEG signal at time $k$ and by $x[k]\in \mathbb{R}^n$ the vector encapsulating all $n$ EEG signals at time $k$. To construct a dynamic model that governs the evolution of the $j$-th EEG signal over a time window, we proceed as follows: (\textit{i}) A fractional order derivative $\Delta^{\alpha_{j}}x_{j}[k]$ captures both short-range memory properties (ie, $\alpha_{j} = 1$) and long-range memory properties (ie, $0 < \alpha_{j} < 1$) of the $j$-th signal. (\textit{ii}) To model the interdependencies between all $n$ signals, as well as the effect of external perturbations, the rate of (memory-based) change of the signal $x_{j}[k]$ can be expressed as $\Delta^{\alpha_{j}}x_{j}[k] = \sum_{i=1}^{n} a_{ji}x_{i}[k] + \sum_{i=1}^{n} b_{i}u_{i}[k]$. Using fractional dynamical concepts~\cite{Gupta2018DealingWU}, a compact dynamical network representation of the brain activity can be expressed as follows:
\begin{equation}
\label{model-equation}
    \begin{aligned}
        \Delta^{\alpha} [k+1] &= Ax[k] + Bu[k] \\
        y[k] &= Cx[k]
    \end{aligned}
\end{equation}
where, $ x \in \mathbb{R}^n$ denotes the state of the brain in action that performs the cognitive tasks, $ u \in \mathbb{R}^p $ corresponds to unobserved external perturbations (e.g., sudden sounds, music), and $ y \in \mathbb{R}^n $ represents the measurable outputs. 

This dynamical network formalism offers several advantages in modeling the brain activity: The parameter of the fractional derivative allows for flexible capture of both short-term fluctuations and long-term memory by adjusting the order of the derivative operator $ \Delta^{\alpha}$ across all signals. When the order is an integer, such as $ \alpha = 1 $, the model reduces to a first-order difference equation, e.g., $ \Delta^1 x[k] = x[k] - x[k - 1] $, characteristic of memoryless or one-step memory systems. In contrast, fractional-order dynamics enable the incorporation of extensive historical information, which is critical for modeling EEG signals that exhibit long-range dependencies. Moreover, the fractional-order operator for the $j$-th state ($ 1 \leq j \leq n $) is expanded as follows:
\begin{equation}
\label{expantion}
    \Delta^{\alpha_{j}} x_{j}[k] = \sum_{i=0}^{k} \psi(\alpha_{j}, i) x_{j}[k - i]
\end{equation}
where $\alpha_{j}$ is the fractional order of $j$-th state, 
$\psi(\alpha_{j}, i) = \frac{\Gamma(i - \alpha_{j})}{\Gamma(-\alpha_{j}) \Gamma(i+1)}$ and $\Gamma(\cdot)$ is the gamma function. Equation (\ref{expantion}) captures the long-term memory effects by weighting all past states $ x_{j}[k - i]$, enabling the system to reflect cumulative physiological changes over time. The dynamics of the brain during a cognitive task are captured by the tuple $ (\boldsymbol{\alpha}, A, B, C) $, where $A$ encodes intrinsic couplings among physiological variables, and $ B $ maps the impact of latent external perturbations, under the assumption $ p < n $. The matrix $A$ captures correlations between internal states, which may vary across fatigue levels and gives us some insight into underlying dependencies. To calculate this coupling relationship or matrices, we used a latent input estimation framework that accounts for unknown input influences as well. This framework incorporates unmeasured inputs as latent variables. These inputs are inferred through the system dynamics within a fractional-order linear framework. The model parameters are estimated using an expectation-maximization algorithm, enabling reliable inference under noisy and sparse conditions.

In a nutshell, the cognition of the brain can be understood as a set of interdependent processes (captured by coupling matrix $A$ in eq. (\ref{model-equation})) mining the acquired data to extract knowledge and exploiting thought, experience, and senses to make decisions. One simple example of a cognitive output could be represented by the emerging thought and phrasing of a sentence. With this analogy in mind and considering the trajectory of coupling matrices as a high-dimensional set of creative sentences, one can ask how we can quantify the complexity and how this complexity changes from high cognitive clarity to cognitive fatigue. Given that the coupling matrix $A$ captures the strengths and patterns of interaction between EEG signals, we flattened and temporarily concatenated them into a high-dimensional set of trajectories for which we computed the Lempel-Ziv complexity index (LZCI)~\cite{ziv1977universal}. The LZCI analyzes the trajectories from left to right, identifying novel patterns, quantifying information gain (counting the number of distinct substrings within a string), and measuring the complexity of the trajectories as the magnitude of diversity of patterns.
Figure~\ref{fig:ci_dq}(g-h) shows the LZCI for a monfractal system (g), the density of the LZCI during the various cognitive levels (h) and the violin plot of LZCI for the various cognitive levels and the monfractal system (i). Of note, the higher the LZCI is, the higher the quantity and frequency of unexpected patterns within the dynamic network evolution, suggesting a higher degree of diversity!\cite{abásolo_simons_morgado_da_silva_tononi_vyazovskiy_2015,höhn_hahn_lendner_hoedlmoser_2024}. Conversely, lower LZCI values suggest a small prevalence of unexpected patterns~\cite{abásolo_simons_morgado_da_silva_tononi_vyazovskiy_2015,höhn_hahn_lendner_hoedlmoser_2024}. While from Figures~\ref{fig:ci_dq}(h) and \ref{fig:ci_dq}(i) we can observe differences between various levels of cognitive fatigue, we also compare the LZCI profile with that of a monofractal system in \ref{fig:ci_dq}(g). The LZCI values for the monofractal case are skewed to the left, with most of the support of the density below 0.9, indicating that its complexity is lower and contains a large amount of expected (similar) patterns. In contrast, the average LZCI values for cognitive levels 0, 1 and 2 are 1.1703, 1.2142, and 1.2320, respectively. This indicates an increasing complexity trend with the transition from cognitive clarity to a higher degree of fatigue. We tested the statistical significance of these results using the Kruskal-Wallis test, which indicates a $p$-value of 0.0671. This analysis suggests that the LZCI values act as a quantitative marker to distinguish between different levels of cognitive fatigue, as it gradually increases from level 0 to 2. We also observe a positive correlation between the higher number of distinct patterns found and the LZCI values of the sequence of dynamic networks.

\vspace{2em}
\noindent\textbf{A non-Markovian dynamical network-based learning approach can predict the emergence of cognitive fatigue with high precision, unlike traditional machine learning approaches that ignore multifractal and network patterns.} We exploit the above-mentioned fractional dynamical network framework and develop a two-step contrastive learning framework followed by a classification layer to use both raw EEG signals and extracted coupling matrices. The first step of our new deep learning (DL) architecture consists of two parallel CNN encoders that process each input separately. During the pre-training phase, these encoders learn to project both data types into a shared latent space using a contrastive loss function, which maximizes agreement between matching pairs while minimizing similarity between non-matching pairs. The second stage builds upon these aligned representations to predict cognitive fatigue levels. This classification component uses two convolution layers that process the aligned features from the first stage. The DL network incorporates batch normalization between convolutional layers to stabilize learning dynamics, with a final softmax activation function that outputs probabilities across three distinct fatigue levels. To avoid overfitting, we used early stopping, a learning rate scheduler, and dropout layers in the neural network design. For classification, the pre-trained encoders are combined with a downstream classifier. The latent representations from both modalities are concatenated and fed into a softmax-activated dense layer that predicts the target classes.  

Our validation methodology incorporates multiple strategies to ensure robustness and generalizability of the model. We implement a $k$-fold cross-validation strategy and vary the $k$ parameter ($k = 5, 8, 10$) by randomly partitioning the dataset into $k$ equal segments, iteratively using each segment for testing while the remaining data serves for training. 
% Additionally, we used leave-one-out validation, where individual participant data is sequentially excluded from training and used exclusively for evaluation. 

% The results of this validation are provided in our supplementary material \hl{supp link}. 

%\subsection{Fractional Dynamics Deep Learning Predicting Cognitive Fatigue Levels }

\begin{figure*}
    \centering
    \includegraphics[width=\linewidth]{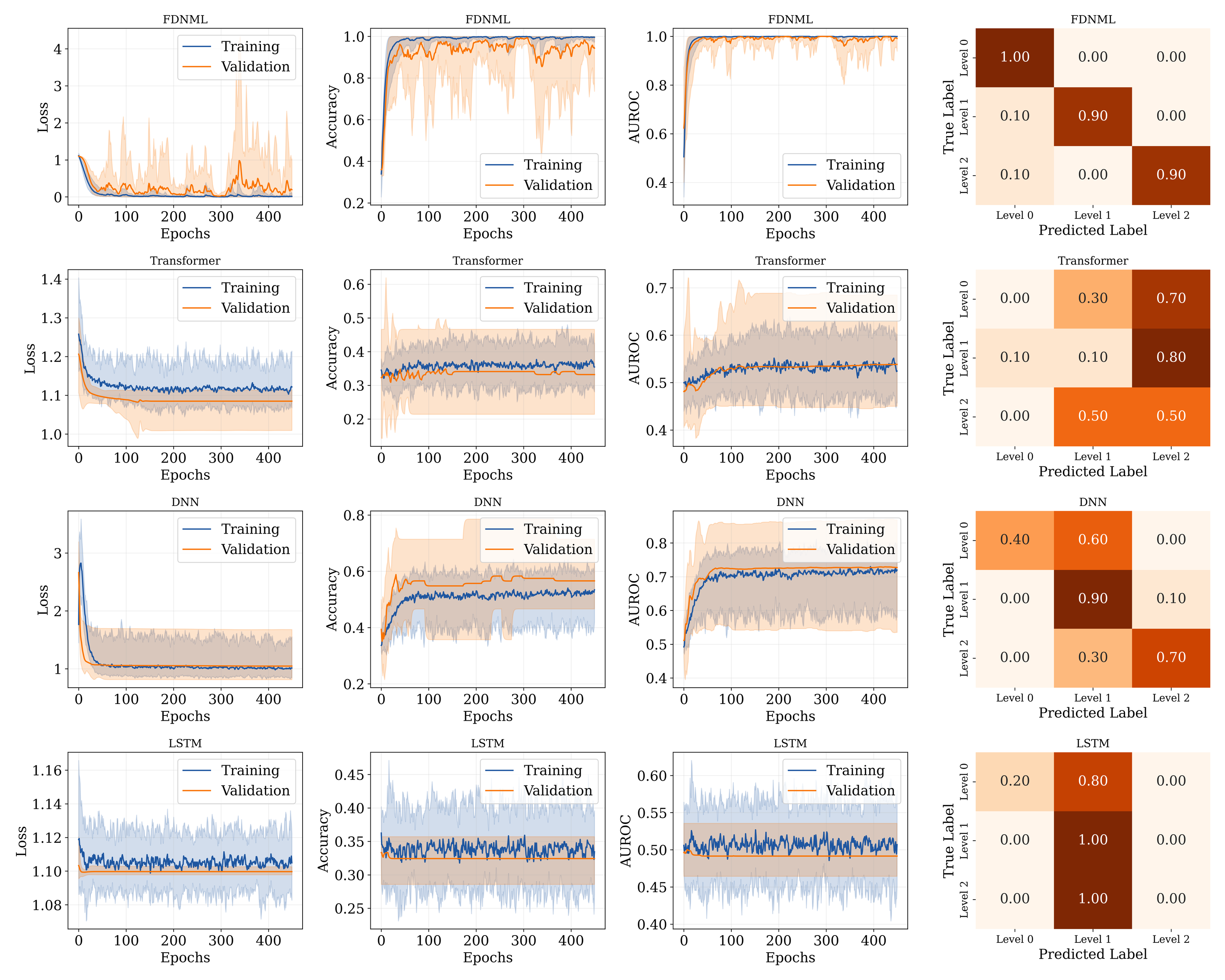}
\caption{\textbf{Comparative training and validation performance of our model and other neural network-based models on EEG data} (a–d) Training and validation loss, accuracy, AUROC, and confusion matrix on unseen data of our model, respectively. (d–h) Training and validation loss, accuracy, AUROC, and confusion matrix of Transformer (i-l). Training and validation loss, accuracy, AUROC, and confusion matrix of DNN (m-p) Training and validation loss, accuracy, AUROC, and confusion matrix of LSTM}
    \label{fig:enter-label}
\end{figure*}

To assess the performance of our proposed machine learning (ML) framework as well as other ML techniques, we performed a cross-validation and first processed all EEG data from the three fatigue levels by learning their fractional dynamic networks. Next, we trained our two-stage ML model on these identified fractional dynamic networks. Our dataset distribution across different levels of fatigue was imbalanced (i.e., instances of non-fatigue were more common than any level of fatigue). We evaluated our results through stratified $k$-fold cross-validation, each fold maintains the same class distribution as the original dataset, leading to a more reliable performance evaluation. We assess the training and validation using the loss, accuracy, F1-score, AUROC, Specificity, Sensitivity, and Precision metrics. We perform a grid search across different hyperparameters, including the number of folds, learning rate, weight decay, and number of epochs. 

Figure~\ref{fig:enter-label} shows the loss, accuracy, and AUROC for the proposed fractional dynamic network-based ML (FDNML) model, a transformer, a deep neural network (DNN), and a long short-term memory (LSTM) approach for both training and validation stages. While all ML models that ignore the multifractal, non-Markovian and network properties of EEG signals show slower decay of the loss and high variability, the loss of the FDNML model decreases fast and exhibits small variation in validation. Similarly, the accuracy of all ML models that ignore multifractal and network patterns ranges from 40\% (for the LSTM) to 60\% (for the DNN), respectively. In contrast, the FDNML model achieves an accuracy of 93\%. A similar trend is also observed when comparing the ML models in terms of AUROC. In conclusion, the FDNML model performed the best without overfitting on 5-fold validation with \textit{AdamW} optimizer with an initial learning rate of $1e-3$ and weight decay of $1e-5$. For completeness, Table~\ref{tab:performance_metrics} summarizes the precision, sensitivity, speciﬁcity, and F1-score of the FDNML model’s prediction results. We conclude that from all these results, the FDNML model exhibits a substantial accuracy except for Precision of level 0, which is 83.33\%.

% \begin{figure*}
%     \centering
%     \includegraphics[width=\linewidth]{Figures/ourmodel.png}
%     \caption{\textbf{Overview of our model training and validation.} Figures \textbf{a-c} show the training and validation accuracy, AUROC, and loss for 450 epochs using 5-fold cross-validation on the CogBeacon dataset. (f) present the heatmap of the confusion matrix of the trained model on the unseen test set with an accuracy of 93\%.}
%     \label{fig:ourmodel}
% \end{figure*}

%Figure~\ref{fig:ourmodel} presents our model training and validation results. Figure~\ref{fig:ourmodel}a shows the training and test results observed from training our model. We observe that training and test accuracies and AUROC increase, while loss decreases during training and validation.

% Classification Training - Time: 357.93s, RAM Usage: 6977.55MB, Trainable Params: 6435

\begin{table}[h!]
\centering
\caption{ Cognitive Fatigue Level predicting results for test set with our model}
\begin{tabular}{lcccc}
\hline
\textbf{Metric}       & \textbf{Level 0} & \textbf{Level 1} & \textbf{Level 2} & \textbf{Overall} \\
\hline
Precision             & 83.33\%          & 100.00\%         & 100.00\%         & 94.44\%          \\
Sensitivity  & 100.00\%         & 90.00\%          & 90.00\%          & 93.33\%          \\
Specificity            & 90.00\%          & 100.00\%         & 100.00\%         & 96.67\%          \\
F1-Score               & 90.91\%          & 94.74\%          & 94.74\%          & 93.46\%          \\

\hline
\end{tabular}
\label{tab:performance_metrics}
\end{table}

\begin{table*}[ht]
\centering
\caption{Comparison of machine learning models, their complexity and performance compared to our model.}
\resizebox{\textwidth}{!}{ 
\begin{tabular}{l c c c c c c c c} 
 \hline
Model & RAM(MB)$\downarrow$ & Time(s)$\downarrow$  & \# Parameters$\downarrow$  & Loss$\downarrow$  & Acc(\%)$\uparrow$  & AUROC(\%)$\uparrow$ & Sens(\%)$\uparrow$ & Pr(\%)$\uparrow$ \\
 \hline
LSTM         & 19.80     & 3526.50 & $43.27\text{K}$   & 1.04& 40.00 & 56.44 & 40.00 & 45.24 \\
KNN          & 0.07    & 0.01   & NA       & -      & 56.67 & 83.50 & 56.67 & 56.34 \\
SVM          & 7.02    & 0.60   & NA       & 0.22& 90.00 & 95.67 & 90.00 & 91.11 \\
DNN          & 7.28  & 128.93  &  476\text{K} $<$ & 0.77& 66.67 & 83.00 & 66.67 & 69.24 \\
Transformer  & 55.03  & 326.41  & 19.9\text{M} $<$ & 2.55& 76.67 & 85.08 & 76.67 & 78.07 \\
FDNML         & 42.21 &52.89      & \textbf{6.24$\text{K}$ }   & 0.91& \textbf{93.33} &\textbf{ 95.00} & \textbf{93.33} & \textbf{94.44} \\
 \hline
\end{tabular}

}
\vspace{1em}
\parbox{0.99\linewidth}{
        \small
        \centering
        Acc: Accuracy, Sens: Sensitivity, Pr: Precision
        }
\label{raw_transposed}

\end{table*}

\vspace{2em}
\noindent\textbf{Model Evaluation and Comparison.}
To benchmark our model against both conventional time series classifiers (e.g., SVM, KNN, LSTM) and recent architectures such as Transformers, we trained each model under identical settings using combined EEG and coupling matrix inputs, and compared performance across metrics including accuracy, AUROC, sensitivity, and specificity, as well as computational demands such as trainable parameters, runtime, and memory usage (see Methods for implementation details). As shown in Table~\ref{raw_transposed}, FDNML surpasses traditional models-DNN (66.67\% accuracy, 83.00\% AUROC), SVM (90.00\%, 95.67\%), KNN (56.67\%, 83.50\%), and LSTM (40.00\%,56.44\%)—as well as a Transformer-based model (76.67\%, 85.08\%). To isolate the contribution of coupling matrices, we also trained baseline models using only raw EEG inputs, applying 8-fold stratified cross-validation  (Table~\ref{raw_transposed}). RAM usage and execution time are computed on the 80\% train dataset for KNN and SVM and as an average for neural networks. The KNN algorithm does not optimize any loss; there is a hinge loss for SVM and a categorical cross-entropy loss for neural networks. As we can see, SVM shows the highest test accuracy among the baselines. However, unlike neural networks, SVM was not trained during cross-validation before applying it to the test set, so there is a high chance of overfitting \cite{svmguide}. Other models failed to achieve 90 \% test accuracy. KNN is not reliable in general; being non-parametric, the algorithm depends highly on data and tends to overfit \cite{halder2024knn}.

% on 80\% of the data while holding out 20\% as a test set

\begin{figure}
    \centering
    \includegraphics[width=\linewidth]{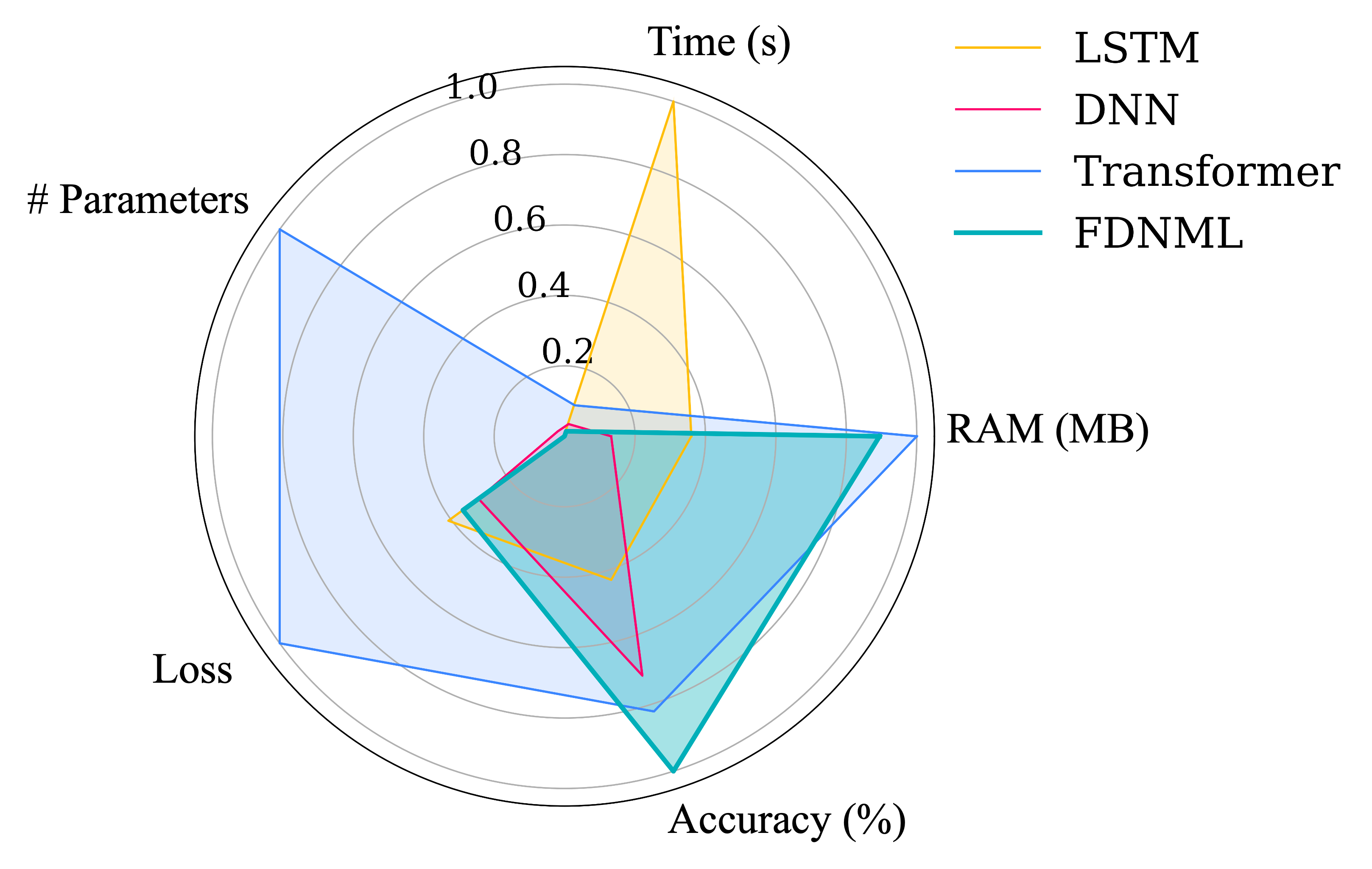}
    \caption{\textbf{The radar plot of comparing the complexity and prediction performance of different deep learning models under k-fold validation} (k = 8): FDNML, Deep neural network (DNN), long short-term memory (LSTM), and Transformer. We normalized all the values represented in this plot.}
    \label{fig:enter-label}
\end{figure}
We present average cross-validation training and validation loss and confusion matrices for neural networks in Figure \ref{fig:enter-label}. As we can see, Transformer may be too complicated for our relatively small dataset. For the same reason, DNN shows high variance in the validation dataset. LSTM is not learned at all on raw EEG data. The confusion matrices show that all the baseline models are unreliable.

\section{Discussion}

% In this study, we proposed the FDNML framework that detects cognitive fatigue across three levels. Based on the multifractal properties of cognitive fatigue across different stages, we modeled the EEG data and generated the Fractional order dynamical networks of each data and fatigue level. We used the dynamical network and the raw EEG data in a parallel pipeline to detect Cognitive fatigue. The limitations of our work including testing the framework on a signal data...

Mental fatigue manifests as a gradual difficulty concentrating on specific tasks, an increased number of mistakes, slower vigilance, response, and processing speed, as well as a general feeling of mental exhaustion that occurs over time due to sustained mental effort and independent of sleepiness. Although we know that failures in early detection of cognitive fatigue contribute to catastrophic events, we still lack rigorous, accurate, and interpretable mathematical understanding, modeling, detection, and forecasting of the onset of various levels of cognitive fatigue. To fill this knowledge gap, we introduced an interpretable fractional dynamical networks-based machine learning (FDNML) framework for real-time detection and characterization of cognitive fatigue. By modeling EEG signals through coupled fractional-order differential equations, our method captures the non-Markovian and time-varying interdependencies inherent in brain activity, which are properties often overlooked by traditional approaches. Our analysis indicates that the multifractal features of brain signals shift across different levels of fatigue, and the complexity of their interdependence also varies in meaningful ways. These findings not only enable us to track fatigue as a continuous process but also support accurate classification of fatigue levels, achieving 93.33\% accuracy and 95\% AUROC. All these findings are based on the most available but a small set of EEG channels. For a more comprehensive understanding of cognitive fatigue and how our approach could be applied in practice for real-time cognitive fatigue assessment, we foresee the following research problems to be further explored: With the increasing ability of rich multimodal sensing of brain activity and physiological processes (e.g., eye movement, heart rate, oxygenation levels, blood glucose variability, stress levels), one can extend our framework to identify the minimum number of multimodal signals and the minimum number of samples required to detect cognitive fatigue transitions in real-time.

 Equally important, our interpretable mathematical modeling and machine learning framework allows for investigating the effect of external perturbations on delaying the onset of dangerous cognitive fatigue levels or even improving the cognitive performance for a finite period of time. For instance, in a human-AI collaborative environment, like a human pilot guiding a drone or swarm of drones across a geographical region or an airplane cockpit, one can exploit our FDNML framework to determine which external perturbation (e.g., playing preferred music, reminding about a family event) improved the attention and cognitive performance for a period of time that exceeds the critical time of landing the airplane. Moreover, we are now learning that the way AI presents complex and conflicting information to humans can impact their decisions and cognitive fatigue transition~\cite{Asgari2023ImpactOE,10.1145/3746059.3747647,Widhanarto2025BeyondST} we foresee that one can exploit our FDNML framework to determine which type of information presentation in optimal by an AI agent to specific individual in order to minimize the likelihood of the onset of various cognitive fatigue levels or obtain the most ethical decision in life-threatening situations. \\

\textbf{Comparison with prior work.} Some EEG-based fatigue studies have achieved high accuracies, but often with complex architectures or datasets that are indirectly related to true cognitive fatigue. For example, Nooh et al.~\cite{Nooh2025An} combined multi-modal biosignals with BOOA feature selection, a graph convolutional autoencoder, and MOHO-based hyperparameter tuning to reach 97.59\% binary accuracy on the MEFAR dataset, at the cost of a highly involved training pipeline and model structure. Similar trends appear in driving-fatigue and construction-safety work: Li et al.’s \textit{E‑5‑D‑BCDRNet} fuses 3D brain power maps with regional rhythm features to detect driving brain fatigue with 88.5\% accuracy~\cite{Li2023An}, while Wang et al.~\cite{Wang2021EEG} encode PDC-based brain functional networks and graph metrics to obtain 87.16\% accuracy. Deep learning approaches, such as Bi-LSTM and CNN models, can further enhance performance, for example, achieving 99.94\% three-class accuracy for construction operators’ mental fatigue~\cite{Mehmood2023Deep} or 88.85\% for construction workers’ cognitive fatigue~\cite{Wang2023Identifying}. However, these models typically involve numerous parameters, lengthy training times, and large datasets. Other work focuses on multi-domain or connectivity features to classify "fatigue" or SEED‑VIG vigilance levels, achieving 87.10\% three-class SEED‑VIG accuracy and 97.40\% two-class accuracy on self-collected data~\cite{Chen2022EEG-based}, or up to 96.57\% three-class accuracy on SEED‑VIG and 99.23\% on the Mendeley driving dataset with a CNN–LSTM on complex network and frequency–spatial features~\cite{Wang2025EEG-based}. 

Compared with graph-based CNN–LSTM pipelines~\cite{Wang2025EEG-based}, generative–adversarial transfer networks~\cite{Zeng2021An}, or multi-branch spatio–temporal networks for fatigue driving~\cite{Gao2023CSF-GTNet:,Li2023An}, the proposed model achieves competitive multi-class performance without relying on heavy fusion modules or very deep networks. Unlike Bi‑LSTM approaches that require millions of samples and complex temporal modeling~\cite{Mehmood2023Deep}, the current design demonstrates that careful feature design and lightweight modeling can deliver high accuracy on rigorously defined fatigue labels. Our method is a practical option for real-world cognitive fatigue monitoring, balancing accuracy, reliable labels, and manageable computational demands.

\textbf{Relevance to clinical Research.} Cognitive fatigue is a common and disabling symptom across neurological and post-infectious conditions, impairing attention, processing speed, decision-making, and daily functioning, yet it often goes undetected in standard clinical assessments \cite{Kunasegaran2023MentalFatigue,Linnhoff2019MSFatigue,Johansson2021MTBI,Ceban2021PostCOVID}.
It reduces quality of life and work capacity in conditions such as multiple sclerosis~\cite{Linnhoff2019MSFatigue}, mild traumatic brain injury (mTBI)~\cite{Golan2018SubjectiveFatigue}, ME/CFS, and post-COVID-19 syndrome. Many patients report lasting cognitive slowing and mental effort, even when standard neuropsychological tests or brain imaging appear normal \cite{Johansson2021MTBI,Ceban2021PostCOVID}.

After COVID-19 alone, persistent fatigue and cognitive problems affect a substantial minority of survivors. These symptoms are linked to inflammatory changes and significant functional limitations\cite{Ceban2021PostCOVID}. At the same time, subjective fatigue ratings show only modest or inconsistent associations with conventional neuropsychological performance. This highlights a clear gap between what patients feel and what current clinical tools can measure \cite{Linnhoff2019MSFatigue,Golan2018SubjectiveFatigue,MatiasGuiu2022PostCOVIDPredictors}.This gap has driven growing research on objective and sensitive markers of cognitive fatigue, including eye-tracking, EEG-based network metrics, signal detection theory measures, speech analysis, and AI-driven biosignal or digital phenotyping approaches to detect performance decline and underlying neural changes during sustained cognitive demand \cite{Yamada2018EyeTracking,Lou2025EEGFatigue,Wylie2021SDT,Nooh2025AIFatigue,Han2024Bibliometric,Dias2024SpeechFatigue,Roman2022SDTMS}. 

Early and accurate detection of cognitive fatigue can reduce safety-critical errors, guide rest and workload decisions, and improve how treatments are evaluated. As a result, better cognitive fatigue assessment is now a major unmet need in both clinical care and research \cite{Karim2024Survey,Linnhoff2019MSFatigue,Walker2019Interventions,Han2024Bibliometric,Dias2024SpeechFatigue,Roman2022SDTMS}.

\textbf{Data Availability: } The data that was used in this study is openly available in \href{https://github.com/MikeMpapa/CogBeaconMultiModal_Dataset_for_Cognitive_Fatigue.git}{Github repository} and cited as~\cite{Papakostas2019CogBeaconAM}. Our findings, features obtained form data and methodology implementation codes are available at our GitHub repository \href{https://github.com/aceatusc/CognitiveFatigueDetection.git}{here}.

\section*{References}

\bibliographystyle{IEEEtran}
\bibliography{sn-bibliography}

@article{leonarduzzi2016p,
  title={p-exponent and p-leaders, Part II: Multifractal analysis. Relations to detrended fluctuation analysis},
  author={Leonarduzzi, Roberto and Wendt, Herwig and Abry, Patrice and Jaffard, St{\'e}phane and Melot, Clothilde and Roux, St{\'e}phane G and Torres, Mar{\'\i}a E},
  journal={Physica A: Statistical Mechanics and its Applications},
  volume={448},
  pages={319--339},
  year={2016},
  publisher={Elsevier}
}

@article{Nooh2025An,title={An exploratory analysis of longitudinal artificial intelligence for cognitive fatigue detection using neurophysiological based biosignal data},author={S. Nooh and Mahmoud Ragab and Rania Aboalela and A. A. Al-Ghamdi and O. Abdulkader and Ghadah Alghamdi},journal={Scientific Reports},year={2025},volume={15},doi={10.1038/s41598-025-96816-8}}

@article{Gao2023CSF-GTNet:,title={CSF-GTNet: A Novel Multi-Dimensional Feature Fusion Network Based on Convnext-GeLU- BiLSTM for EEG-Signals-Enabled Fatigue Driving Detection},author={Dongrui Gao and Pengrui Li and Manqing Wang and Yujie Liang and Shihong Liu and Jiliu Zhou and Lutao Wang and Yongqing Zhang},journal={IEEE Journal of Biomedical and Health Informatics},year={2023},volume={28},pages={2558-2568},doi={10.1109/jbhi.2023.3240891}}

@article{Zeng2021An,title={An EEG-Based Transfer Learning Method for Cross-Subject Fatigue Mental State Prediction},author={Hong Zeng and Xiufeng Li and G. Borghini and Yue Zhao and P. Aricó and G. D. Flumeri and Nicolina Sciaraffa and Wael Zakaria and Wanzeng Kong and F. Babiloni},journal={Sensors (Basel, Switzerland)},year={2021},volume={21},doi={10.3390/s21072369}}

@article{Wang2025EEG-based,title={EEG-based fatigue state evaluation by combining complex network and frequency-spatial features},author={Kefa Wang and Xiaoqian Mao and Yuebin Song and Qiuyu Chen},journal={Journal of Neuroscience Methods},year={2025},volume={416},doi={10.1016/j.jneumeth.2025.110385}}

@article{Wang2023Identifying,title={Identifying mental fatigue of construction workers using EEG and deep learning},author={Yao Wang and Yuecheng Huang and bo-yi gu and Sihan Cao and D. Fang},journal={Automation in Construction},year={2023},doi={10.1016/j.autcon.2023.104887}}

@article{Chen2022EEG-based,title={EEG-based mental fatigue detection using linear prediction cepstral coefficients and Riemann spatial covariance matrix},author={Kun Chen and Zhiyong Liu and Quan Liu and Qingsong Ai and Li Ma},journal={Journal of Neural Engineering},year={2022},volume={19},doi={10.1088/1741-2552/aca1e2}}

@article{Mehmood2023Deep,title={Deep learning-based construction equipment operators' mental fatigue classification using wearable EEG sensor data},author={Imran Mehmood and Heng Li and Yazan K. Qarout and Waleed Umer and S. Anwer and Haitao Wu and Mudasir Hussain and M. Antwi-Afari},journal={Adv. Eng. Informatics},year={2023},volume={56},pages={101978},doi={10.1016/j.aei.2023.101978}}

@article{Wang2021EEG,title={EEG Driving Fatigue Detection With PDC-Based Brain Functional Network},author={Fei Wang and Shichao Wu and Jingyu Ping and Zongfeng Xu and Hao Chu},journal={IEEE Sensors Journal},year={2021},volume={21},pages={10811-10823},doi={10.1109/jsen.2021.3058658}}

@article{Li2023An,title={An EEG-based Brain Cognitive Dynamic Recognition Network for representations of brain fatigue},author={Pengrui Li and Yongqing Zhang and Shihong Liu and Liqi Lin and Haokai Zhang and Tian Tang and Dongrui Gao},journal={Appl. Soft Comput.},year={2023},volume={146},pages={110613},doi={10.1016/j.asoc.2023.110613}}

@article{jaffard2016p,
  title={p-exponent and p-leaders, Part I: Negative pointwise regularity},
  author={Jaffard, Stephane and Melot, Clothilde and Leonarduzzi, Roberto and Wendt, Herwig and Abry, Patrice and Roux, St{\'e}phane G and Torres, Mar{\'\i}a Eugenia},
  journal={Physica A: Statistical Mechanics and its Applications},
  volume={448},
  pages={300--318},
  year={2016},
  publisher={Elsevier}
}

@article{Mitler1988CatastrophesSA,
  title={Catastrophes, sleep, and public policy: consensus report.},
  author={Merrill M. Mitler and Mary A. Carskadon and Charles A. Czeisler and William C. Dement and David F. Dinges and R. Curtis Graeber},
  journal={Sleep},
  year={1988},
  volume={11 1},
  pages={
          100-9
        },
  url={https://api.semanticscholar.org/CorpusID:26727860}
}

@article{ghorbani2013cyber,
  title={A cyber-physical system approach to artificial pancreas design},
  author={Mahboobeh Ghorbani and Paul Bogdan},
  journal={2013 International Conference on Hardware/Software Codesign and System Synthesis (CODES+ISSS)},
  year={2013},
  pages={1-10},
  url={https://api.semanticscholar.org/CorpusID:11541129}
}

@article{goldberger2002fractal,
  title={Fractal dynamics in physiology: alterations with disease and aging},
  author={Goldberger, Ary L and Amaral, Luis AN and Hausdorff, Jeffrey M and Ivanov, Plamen Ch and Peng, C-K and Stanley, H Eugene},
  journal={Proceedings of the national academy of sciences},
  volume={99},
  number={suppl\_1},
  pages={2466--2472},
  year={2002},
  publisher={National Academy of Sciences}
}

@article{ivanov1999multifractality,
  title={Multifractality in human heartbeat dynamics},
  author={Ivanov, Plamen Ch and Amaral, Luis A Nunes and Goldberger, Ary L and Havlin, Shlomo and Rosenblum, Michael G and Struzik, Zbigniew R and Stanley, H Eugene},
  journal={Nature},
  volume={399},
  number={6735},
  pages={461--465},
  year={1999},
  publisher={Nature Publishing Group UK London}
}

@article{techera2016causes,
  title={Causes and consequences of occupational fatigue: meta-analysis and systems model},
  author={Techera, Ulises and Hallowell, Matthew and Stambaugh, Nathan and Littlejohn, Ray},
  journal={Journal of occupational and environmental medicine},
  volume={58},
  number={10},
  pages={961--973},
  year={2016},
  publisher={LWW}
}

@article{harrington2001health,
  title={Health effects of shift work and extended hours of work},
  author={Harrington, J Malcolm},
  journal={Occupational and Environmental medicine},
  volume={58},
  number={1},
  pages={68--72},
  year={2001},
  publisher={BMJ Publishing Group Ltd}
}

@article{wang2016compensatory,
  title={Compensatory neural activity in response to cognitive fatigue},
  author={Wang, Chao and Trongnetrpunya, Amy and Samuel, Immanuel Babu Henry and Ding, Mingzhou and Kluger, Benzi M},
  journal={Journal of neuroscience},
  volume={36},
  number={14},
  pages={3919--3924},
  year={2016},
  publisher={Society for Neuroscience}
}

@article{borragan2017cognitive,
  title={Cognitive fatigue: A time-based resource-sharing account},
  author={Borrag{\'a}n, Guillermo and Slama, Hichem and Bartolomei, Mario and Peigneux, Philippe},
  journal={Cortex},
  volume={89},
  pages={71--84},
  year={2017},
  publisher={Elsevier}
}

@article{wiehler2022neuro,
  title={A neuro-metabolic account of why daylong cognitive work alters the control of economic decisions},
  author={Wiehler, Antonius and Branzoli, Francesca and Adanyeguh, Isaac and Mochel, Fanny and Pessiglione, Mathias},
  journal={Current Biology},
  volume={32},
  number={16},
  pages={3564--3575},
  year={2022},
  publisher={Elsevier}
}

@article{ziv1977universal,
  title={A universal algorithm for sequential data compression},
  author={Ziv, Jacob and Lempel, Abraham},
  journal={IEEE Transactions on information theory},
  volume={23},
  number={3},
  pages={337--343},
  year={1977},
  publisher={IEEE}
}

@book{doukhan2002theory,
  title={Theory and applications of long-range dependence},
  author={Doukhan, Paul and Oppenheim, George and Taqqu, Murad},
  year={2002},
  publisher={Springer Science \& Business Media}
}

@article{Peng2002QuantifyingFD,
  title={Quantifying Fractal Dynamics of Human Respiration: Age and Gender Effects},
  author={Chung-Kang Peng and Joseph E. Mietus and Yanhui Liu and Christine Lee and Jeffrey M. Hausdorff and Harry Eugene Stanley and Ary L. Goldberger and Lewis A. Lipsitz},
  journal={Annals of Biomedical Engineering},
  year={2002},
  volume={30},
  pages={683-692},
  url={https://api.semanticscholar.org/CorpusID:13099492}
}

@article{Schumann2010AgingEO,
  title={Aging effects on cardiac and respiratory dynamics in healthy subjects across sleep stages.},
  author={Aicko Y. Schumann and Ronny P. Bartsch and Thomas Penzel and Plamen Ch. Ivanov and Jan W. Kantelhardt},
  journal={Sleep},
  year={2010},
  volume={33 7},
  pages={
          943-55
        },
  url={https://api.semanticscholar.org/CorpusID:34574292}
}

@article{karim2024examining,
  title={Examining the landscape of cognitive fatigue detection: A comprehensive survey},
  author={Karim, Enamul and Pavel, Hamza Reza and Nikanfar, Sama and Hebri, Aref and Roy, Ayon and Nambiappan, Harish Ram and Jaiswal, Ashish and Wylie, Glenn R and Makedon, Fillia},
  journal={Technologies},
  volume={12},
  number={3},
  pages={38},
  year={2024},
  publisher={MDPI}
}

@article{Chang2007FractalDE,
  title={Fractal Dimension Estimation Via Spectral Distribution Function and Its Application to Physiological Signals},
  author={Shyang Chang and Shiun-Jeng Li and Meng-Ju Chiang and Shih-Jen Hu and Ming-Chun Hsyu},
  journal={IEEE Transactions on Biomedical Engineering},
  year={2007},
  volume={54},
  pages={1895-1898},
  url={https://api.semanticscholar.org/CorpusID:8746834}
}

@article{Gupta2018DealingWU,
  title={Dealing with Unknown Unknowns: Identification and Selection of Minimal Sensing for Fractional Dynamics with Unknown Inputs},
  author={Gaurav Gupta and S{\'e}rgio Daniel Pequito and Paul Bogdan},
  journal={2018 Annual American Control Conference (ACC)},
  year={2018},
  pages={2814-2820},
  url={https://api.semanticscholar.org/CorpusID:3845530}
}

@article{Papakostas2019CogBeaconAM,
  title={CogBeacon: A Multi-Modal Dataset and Data-Collection Platform for Modeling Cognitive Fatigue},
  author={Michalis Papakostas and Akilesh Rajavenkatanarayanan and Fillia Makedon},
  journal={Technologies},
  year={2019},
  url={https://api.semanticscholar.org/CorpusID:196185638}
}

@article{cho_lee_2022, 
title={Forecasting the Volatility of the Stock Index with Deep Learning Using Asymmetric Hurst Exponents}, 
volume={6}, 
DOI={https://doi.org/10.3390/fractalfract6070394}, number={7}, journal={Fractal and Fractional}, 
author={Cho, Poongjin and Lee, Minhyuk}, year={2022}, month={Jul}, pages={394} }

@article{gaurav_anand_kumar_2021, 
title={EEG based cognitive task classification using multifractal detrended fluctuation analysis}, volume={15}, DOI={https://doi.org/10.1007/s11571-021-09684-z}, number={6}, journal={Cognitive Neurodynamics}, author={Gaurav, G. and Anand, R. S. and Kumar, Vinod}, year={2021}, month={May} }

@article{ tibarewala_2017, 
title={Fractal analysis of EEG signals for studying the effect of cognitive stress on brain}, volume={25}, DOI={https://doi.org/10.1504/ijbet.2017.087707}, number={2/3/4}, journal={International Journal of Biomedical Engineering and Technology}, publisher={Inderscience Publishers},
author={Rakshit, Arnab and Banerjee, Anwesha and Mazumder, Ankita and Ghosh, Poulami and Dey, Anilesh and D.N. Tibarewala}, year={2017}, month={Jan}, pages={336–336} }

@article {abásolo_simons_morgado_da_silva_tononi_vyazovskiy_2015,
author={Abásolo, Daniel and Simons, Samantha and Morgado da Silva, Rita and Tononi, Giulio and Vyazovskiy, Vladyslav V.},
title={Lempel-Ziv complexity of cortical activity during sleep and waking in rats}, 
volume={113}, 
DOI={https://doi.org/10.1152/jn.00575.2014}, number={7}, journal={Journal of Neurophysiology}, year={2015}, month={Apr}, pages={2742–2752} }

@article{Martins2021Fatigue,title={Fatigue Monitoring Through Wearables: A State-of-the-Art Review},author={Neusa R. Adão Martins and S. Annaheim and C. Spengler and R. Rossi},journal={Frontiers in Physiology},year={2021},volume={12},doi={10.3389/fphys.2021.790292}}

@article{höhn_hahn_lendner_hoedlmoser_2024, 
title={Spectral Slope and Lempel-Ziv complexity as robust markers of brain states during sleep and wakefulness}, volume={11}, DOI={https://doi.org/10.1523/eneuro.0259-23.2024}, number={3}, journal={eNeuro}, publisher={Society for Neuroscience}, author={Höhn, Christopher and Hahn, Michael A and Lendner, Janna D and Hoedlmoser, Kerstin}, year={2024}, month={Mar}, pages={ENEURO.0259-23.2024} }

@article{halder2024knn,
  title   = {Enhancing K-nearest Neighbor Algorithm: A Comprehensive Review and Performance Analysis of Modifications},
  author  = {Rajib Kumar Halder and Mohammed Nasir Uddin and Md. Ashraf Uddin and Sunil Aryal and Ansam Khraisat},
  journal = {Journal of Big Data},
  year    = {2024},
  volume  = {11},
  pages   = {113},
  doi     = {10.1186/s40537-024-00973-y},
  url     = {https://doi.org/10.1186/s40537-024-00973-y}
}

@techreport{svmguide,
  title        = {A Practical Guide to Support Vector Classification},
  author       = {Chih-Wei Hsu and Chih-Chung Chang and Chih-Jen Lin},
  institution  = {Department of Computer Science, National Taiwan University},
  year         = {2016},
  note         = {Initial version 2003; last updated 19 May 2016},
  url          = {http://www.csie.ntu.edu.tw/~cjlin/papers/guide/guide.pdf}
}

@article{karim2023fatigue,
  title={An EEG-based Cognitive Fatigue Detection System},
  author={Enamul Karim and Hamza Reza Pavel and Ashish Jaiswal and Mohammad Zaki Zadeh and Michail Theofanidis and Glenn R. Wylie and Fillia Makedon},
  journal={Proceedings of the 16th International Conference on PErvasive Technologies Related to Assistive Environments},
  year={2023},
  url={https://api.semanticscholar.org/CorpusID:260777077}
}

@article{gao2023logmelcrnn,
  title   = {EEG Driving Fatigue Detection Based on Log-Mel Spectrogram and Convolutional Recurrent Neural Networks},
  author  = {Dongrui Gao and Xue Tang and Manqing Wan and Guo Huang and Yongqing Zhang},
  journal = {Frontiers in Neuroscience},
  year    = {2023},
  volume  = {17},
  pages   = {1136609},
  doi     = {10.3389/fnins.2023.1136609},
  url     = {https://doi.org/10.3389/fnins.2023.1136609}
}

@article{lee2024pilotfatigue,
  title   = {Decoding Fatigue Levels of Pilots Using EEG Signals with Hybrid Deep Neural Networks},
  author  = {Dong-Hyeon Lee and Seong-Jun Kim and Se-Ho Kim},
  journal = {arXiv preprint arXiv:2411.09707},
  year    = {2024},
  url     = {https://arxiv.org/abs/2411.09707}
}

@article{lempel_ziv,
  title   = {On the Complexity of Finite Sequences},
  author  = {Lempel, Abraham and Ziv, Jacob},
  journal = {IEEE Transactions on Information Theory},
  volume  = {22},
  number  = {1},
  pages   = {75--81},
  year    = {1976},
  doi     = {10.1109/TIT.1976.1055501}
}

@article{Kunasegaran2023MentalFatigue,
  author  = {Kunasegaran, K. and Ismail, A. and Ramasamy, S. and Gnanou, J. and Caszo, B. and Chen, P.},
  title   = {Understanding mental fatigue and its detection: a comparative analysis of assessments and tools},
  journal = {PeerJ},
  year    = {2023},
  volume  = {11},
  doi     = {10.7717/peerj.15744}
}

@article{Linnhoff2019MSFatigue,
  author  = {Linnhoff, S. and Fiene, M. and Heinze, H. and Zaehle, T.},
  title   = {Cognitive Fatigue in Multiple Sclerosis: An Objective Approach to Diagnosis and Treatment by Transcranial Electrical Stimulation},
  journal = {Brain Sciences},
  year    = {2019},
  volume  = {9},
  doi     = {10.3390/brainsci9050100}
}

@article{Johansson2021MTBI,
  author  = {Johansson, B.},
  title   = {Mental Fatigue after Mild Traumatic Brain Injury in Relation to Cognitive Tests and Brain Imaging Methods},
  journal = {International Journal of Environmental Research and Public Health},
  year    = {2021},
  volume  = {18},
  doi     = {10.3390/ijerph18115955}
}

@article{Ceban2021PostCOVID,
  author  = {Ceban, F. and Ling, S. and Lui, L. and Lee, Y. and Gill, H. and Teopiz, K. and Rodrigues, N. and Subramaniapillai, M. and Di Vincenzo, J. and Cao, B. and Lin, K. and Mansur, R. and Ho, R. and Rosenblat, J. and Miskowiak, K. and Vinberg, M. and Maleti{\'c}, V. and McIntyre, R.},
  title   = {Fatigue and cognitive impairment in Post-COVID-19 Syndrome: A systematic review and meta-analysis},
  journal = {Brain, Behavior, and Immunity},
  year    = {2021},
  volume  = {101},
  doi     = {10.1016/j.bbi.2021.12.020}
}

@article{Golan2018SubjectiveFatigue,
  author  = {Golan, D. and Doniger, G. and Wissemann, K. and Zarif, M. and Bumstead, B. and Buhse, M. and Fafard, L. and Lavi, I. and Wilken, J. and Gudesblatt, M.},
  title   = {The impact of subjective cognitive fatigue and depression on cognitive function in patients with multiple sclerosis},
  journal = {Multiple Sclerosis Journal},
  year    = {2018},
  volume  = {24},
  doi     = {10.1177/1352458517695470}
}

@article{MatiasGuiu2022PostCOVIDPredictors,
  author  = {Mat{\'i}as-Guiu, J. and Delgado-Alonso, C. and D{\'i}ez-Cirarda, M. and Mart{\'i}nez-Petit, {\'A}. and Oliver-Mas, S. and Delgado-{\'A}lvarez, A. and Cuevas, C. and Valles-Salgado, M. and Gil, M. and Yus, M. and G{\'o}mez-Ruiz, N. and Polidura, C. and Pag{\'a}n, J. and Mat{\'i}as-Guiu, J. and Ayala, J.},
  title   = {Neuropsychological Predictors of Fatigue in Post-COVID Syndrome},
  journal = {Journal of Clinical Medicine},
  year    = {2022},
  volume  = {11},
  doi     = {10.3390/jcm11133886}
}

@article{Lou2025EEGFatigue,
  author  = {Lou, Y. and Pi, R. and Sun, R. and Wu, J. and Wang, W. and Zhu, Z. and Dai, T. and Gong, W.},
  title   = {Graph theory-based analysis of functional connectivity changes in brain networks underlying cognitive fatigue: An EEG study},
  journal = {PLOS One},
  year    = {2025},
  volume  = {20},
  doi     = {10.1371/journal.pone.0329212}
}

@article{Nooh2025AIFatigue,
  author  = {Nooh, S. and Ragab, M. and Aboalela, R. and Al-Ghamdi, A. and Abdulkader, O. and Alghamdi, G.},
  title   = {An exploratory analysis of longitudinal artificial intelligence for cognitive fatigue detection using neurophysiological based biosignal data},
  journal = {Scientific Reports},
  year    = {2025},
  volume  = {15},
  doi     = {10.1038/s41598-025-96816-8}
}

@article{Yamada2018EyeTracking,
  author  = {Yamada, Y. and Kobayashi, M.},
  title   = {Detecting mental fatigue from eye-tracking data gathered while watching video: Evaluation in younger and older adults},
  journal = {Artificial Intelligence in Medicine},
  year    = {2018},
  volume  = {91},
  doi     = {10.1016/j.artmed.2018.06.005}
}

@article{Wylie2021SDT,
  author  = {Wylie, G. and Yao, B. and Sandry, J. and DeLuca, J.},
  title   = {Using Signal Detection Theory to Better Understand Cognitive Fatigue},
  journal = {Frontiers in Psychology},
  year    = {2021},
  volume  = {11},
  doi     = {10.3389/fpsyg.2020.579188}
}

@article{Han2024Bibliometric,
  author  = {Han, J. and Bai, K. and Zhang, C. and Liu, N. and Yang, G. and Shang, Y. and Song, J. and Su, D. and Hao, Y. and Feng, X. and Li, S. and Wang, W.},
  title   = {Objective assessment of cognitive fatigue: a bibliometric analysis},
  journal = {Frontiers in Neuroscience},
  year    = {2024},
  volume  = {18},
  doi     = {10.3389/fnins.2024.1479793}
}

@article{Dias2024SpeechFatigue,
  author  = {Dias, M. and D{\"o}rr, F. and Garthof, S. and Sch{\"a}fer, S. and Elmers, J. and Schwed, L. and Linz, N. and Overell, J. and Hayward-Koennecke, H. and Tr{\"o}ger, J. and K{\"o}nig, A. and Dillenseger, A. and Tackenberg, B. and Ziemssen, T.},
  title   = {Detecting fatigue in multiple sclerosis through automatic speech analysis},
  journal = {Frontiers in Human Neuroscience},
  year    = {2024},
  volume  = {18},
  doi     = {10.3389/fnhum.2024.1449388}
}

@article{Widhanarto2025BeyondST,
  title={Beyond slides: the impact of gamified Web presentations on student cognitive load},
  author={Ghanis Putra Widhanarto and Zamzami Zainuddin and Titi Prihatin and Sunawan Sunawan and Amirul Mukminin and Seftia Kusumawardani and Mulawarman Mulawarman},
  journal={Information and Learning Sciences},
  year={2025},
  url={https://api.semanticscholar.org/CorpusID:277655451}
}

@inproceedings{10.1145/3746059.3747647,
author = {Huang, Run and Zhao, Anna Katherine and Saghi, Zeinabsadat and Sabouri, Sadra and Chattopadhyay, Souti},
title = {Beyond the Page: Enriching Academic Paper Reading with Social Media Discussions},
year = {2025},
isbn = {9798400720376},
publisher = {Association for Computing Machinery},
address = {New York, NY, USA},
url = {https://doi.org/10.1145/3746059.3747647},
doi = {10.1145/3746059.3747647},
booktitle = {Proceedings of the 38th Annual ACM Symposium on User Interface Software and Technology},
articleno = {90},
numpages = {25},
location = {
},
series = {UIST '25}
}

@article{Asgari2023ImpactOE,
  title={Impact of Electronic Health Record Use on Cognitive Load and Burnout Among Clinicians: Narrative Review},
  author={Elham Asgari and Japsimar Kaur and Gani Nuredini and Jasmine Balloch and Andrew M Taylor and Neil J. Sebire and Robert Robinson and Catherine Peters and Shankar Sridharan and Dominic Pimenta},
  journal={JMIR Medical Informatics},
  year={2023},
  volume={12},
  url={https://api.semanticscholar.org/CorpusID:268465775}
}

@article{Roman2022SDTMS,
  author  = {Roman, C. and DeLuca, J. and Yao, B. and Genova, H. and Wylie, G.},
  title   = {Signal Detection Theory as a Novel Tool to Understand Cognitive Fatigue in Individuals With Multiple Sclerosis},
  journal = {Frontiers in Behavioral Neuroscience},
  year    = {2022},
  volume  = {16},
  doi     = {10.3389/fnbeh.2022.828566}
}

@article{Karim2024Survey,
  author  = {Karim, E. and Pavel, H. and Nikanfar, S. and Hebri, A. and Roy, A. and Nambiappan, H. and Jaiswal, A. and Wylie, G. and Makedon, F.},
  title   = {Examining the Landscape of Cognitive Fatigue Detection: A Comprehensive Survey},
  journal = {Technologies},
  year    = {2024},
  doi     = {10.3390/technologies12030038}
}

@article{Walker2019Interventions,
  author  = {Walker, L. and Lindsay-Brown, A. and Berard, J.},
  title   = {Cognitive Fatigability Interventions in Neurological Conditions: A Systematic Review},
  journal = {Neurology and Therapy},
  year    = {2019},
  volume  = {8},
  doi     = {10.1007/s40120-019-00158-3}
}

\end{document}